# Selective Random Structure Search (SRSS): Unbiased Exploration of Polymorphs in Crystals

Jiexi Song [a], Diwei Shi [b,*], Aixian She [c], Chongde Cao [c,*] and Fengyuan Xuan [a,*]

a Suzhou Laboratory, Suzhou 215123, China
b School of Naval Architecture and Maritime, Zhejiang Ocean University, Zhoushan 316022, China
c School of Physical Science and Technology, Northwestern Polytechnical University, Xian 710072, China

Author to whom correspondence should be addressed:
shidiwei@zjou.edu.cn; caocd@nwpu.edu.cn; xuanfy@szlab.ac.cn

## Abstract

Crystal structure prediction has traditionally relied on prototype-based seeding, approaches that often bias sampling toward known low-energy basins and overlook metastable polymorphs with unconventional symmetries. Here, we introduce Selective Random Structure Search (SRSS), a high-throughput, unbiased framework designed to explore the configurational space of crystalline materials across all dimensions. SRSS combines symmetry-constrained random generation with feature-based diversity selection and rapid relaxation and stability evaluation via universal machine-learning interatomic potentials (uMLIPs). Applied to diverse systems, including bulk system SiC and BaPtAs, 2D layered compounds $NbSe_2$, and 1D nanotubes GaN, SRSS successfully recovers known ground states while revealing numerous previously unreported, dynamically stable polymorphs. Notable discoveries include complex cage-like SiC polytypes, low-energy BaPtAs polymorphs beyond experimental records, a semiconducting orthorhombic phase of 2D-$NbSe_2$, and distinct armchair/zigzag GaN nanotubes. Crucially, the entire workflow operates efficiently on standard CPU resources without GPU acceleration, demonstrating that rigorous, hypothesis-free polymorph discovery is accessible even in resource-limited settings. SRSS thus establishes a robust, scalable platform for mapping the full landscape of crystal stability, bridging the gap between exhaustive search and computational feasibility.

## 1. Introduction

Traditional optimization algorithms for crystal structure prediction, ranging from evolutionary strategies to swarm intelligence, have achieved remarkable success in locating ground-state structures with high efficiency.[1-3] However, these methods typically operate under implicit or explicit constraints on the initial and iterative structural pool, often favoring low-energy basins, high-symmetry prototypes, or chemically intuitive motifs.[4-6] As a consequence, their sampling is inherently limited in breadth, and they may systematically overlook metastable polymorphs that reside in sparsely populated regions of configuration space, particularly those stabilized by strict symmetry requirements or unconventional atomic arrangements.[7-8] Such phases, while thermodynamically suboptimal, can be dynamically stable and exhibit unique electronic, optical, or topological properties.[9-10] The challenge thus lies not merely in accelerating energy evaluation, but in ensuring that the structural search itself is comprehensive, unbiased, and grounded in the fundamental principles of crystallography.[11-12]

To this end, we introduce Selective Random Structure Search (SRSS): a simple yet powerful framework designed to achieve exhaustive, crystallography-guided exploration of metastable polymorphs across crystal dimensions. Rather than relying on iterative optimization or prototype-based seeding, SRSS begins by generating a large ensemble of candidate structures that uniformly span compatible space groups (3D), layer groups (2D), or rod groups (1D), as permitted by composition and dimensionality. This ensures that even symmetry-restricted or geometrically exotic configurations, often missed by conventional samplers, are explicitly included from the outset. Subsequent stages employ feature-based diversity selection and rapid relaxation via a universal machine-learning interatomic potential (uMLIP) to distill physically plausible candidates, followed by rigorous thermodynamic (convex-hull distance) and dynamic (phonon stability) filters. Applied to diverse systems including 3D-SiC, 3D-BaPtAs, 2D-$NbSe_2$, and 1D-GaN, SRSS recovers known polymorphs and, more importantly, reveals numerous previously unreported metastable phases that satisfy both energetic accessibility and lattice-dynamical stability. In doing so, SRSS establishes a physics-driven, high-throughput platform for hypothesis-free polymorph discovery.

## 2. Methods

### 2.1 Overall workflow

The goal of selective random structure search (SRSS) workflow is to navigate the vast configuration space by combining high-throughput symmetry constrained random crystal generation with diversity selection and physics-driven filtering, ultimately yielding a compact set of thermodynamically and dynamically favorable structures.

The SRSS proceeds in six stages. First, a repository of symmetry-constrained candidate structures are generated randomly for each composition, evenly distributed under each group, organized by space group (3D), layer group (2D), rod group (1D) using the PyXtal package.[13] Second, each structure is mapped to a numerical feature vector using dimension-specific descriptors (Section 2.2). Third, feature vectors are standardized and, where necessary, reduced in dimension via Principal Component Analysis (PCA) to mitigate the curse of dimensionality.[14-16] Fourth, diversity-oriented selection algorithms K-means or HDBSCAN clustering are applied in feature space to identify a compact subset of representative structures that maximize coverage of the structural landscape.[17-20] Fifth, all selected structures undergo rapid geometry relaxation using a pretrained universal machine-learning interatomic potential (uMLIP).[21-23] Structures that fail to converge, exhibit unphysical features, or lie far above the low-energy region are discarded. Sixth, the retained structures are subjected to dual-stability filtering: thermodynamic stability is assessed via convex-hull analysis of formation energies,[24] and dynamic stability is verified through phonon spectrum calculations, both accelerated by the uMLIP. The final output is a curated set of structures that are both thermodynamically competitive and dynamically stable within the accuracy of the machine-learning potential.

The SRSS method eliminates reliance on prototypes starting from the structural generation phase by employing a uniform distribution based on symmetry, thereby avoiding the omission of unconventional configurations. By integrating a three-pronged strategy of dimensionality reduction, clustering, and machine learning potential functions, the method significantly reduces the number of structures requiring analysis while maintaining broad, physically meaningful coverage of the configuration space. Consequently, it achieves an efficient balance between unbiased exploration and computational cost control.

**2.2 Structural representation and feature construction**

Each raw crystal is treated as a three-dimensional arrangement of atoms with periodic or non-periodic boundary conditions, depending on its dimensionality. For a set of structures, we construct for each structure a feature vector intended to capture the essential geometrical and local chemical characteristics relevant to structural diversity. These features combine physically interpretable geometric quantities with more expressive local-environment descriptors where appropriate. Prior to selection, all features are standardized so that each component has zero mean and unit variance across the dataset. For high-dimensional descriptors, such as those arising from local environment expansions or matrix-based fingerprints, we further apply PCA to obtain reduced feature vectors that retain a prescribed fraction of the total variance (typically 90–95%). This step improves numerical stability and clustering performance without significantly compromising the structural resolution.

For bulk crystals, we employ two types of representation. The first is a set of simple geometric and topological features, including the unit-cell volume, mass density, the total number of atoms, the basic lattice parameters, and the statistical measures of the fractional atomic coordinates, including mean and standard deviation along each crystallographic direction. These quantities provide a fast and physically transparent description of cell size, packing density, and the gross distribution of atoms within the unit cell. The second type consists of more expressive local-environment descriptors based on smooth overlaps of atomic positions (SOAP).[25] In this representation, the neighborhood of each atom is encoded into a rotationally invariant spectrum, which is then aggregated over all atoms in the unit cell to form a structure-level fingerprint. The radial cutoff used in these descriptors is chosen to be sufficiently large to capture the primary coordination shells while remaining compatible with the typical nearest-neighbor distances in crystalline solids.

For 2D layered structures, we tailor the feature construction to emphasize in-plane properties and layer thickness. The basic geometric descriptors include the in-plane cell area, the areal atomic density, the total number of atoms per cell, and the in-plane lattice constants and their aspect ratio. Atomic positions are analyzed primarily in the plane of the layers. We compute the mean and standard deviation of the in-plane fractional coordinates, capturing the spatial distribution of atoms within the layer. Along the direction perpendicular to the plane, we estimate an effective layer thickness from the spread of fractional coordinates scaled by the out-of-plane lattice parameter. In addition, in-plane interatomic distances (maximum and average) are used to characterize the lateral extent of the structure and typical bond lengths. As in the 3D case, these 2D-specific geometric features can be augmented by SOAP descriptors. 1D structures present a particular challenge due to the presence of large vacuum regions in directions. To obtain a robust representation, we currently use SOAP descriptors for 1D dimension.

**2.3 Diversity-oriented selection in feature space**

Our overall strategy is diversity-oriented: instead of prioritizing extremal energies or other scalar properties, we aim to preserve the structural variety present in the original pool while reducing its size to a computationally tractable subset. To this end, we consider two complementary families of methods: (i) centroid-based clustering via *K*-means,[18-19, 26] which assumes an approximately spherical cluster structure and a user-specified target number of representatives, and (ii) density-based clustering via hierarchical density-based spatial clustering of applications with noise (HDBSCAN),[20] which infers both the number and shape of clusters directly from the data and explicitly identifies low-density, *noisy-rare* structures.

The default selection protocol is based on K-means clustering combined with medoid selection. For a group containing candidate structures with feature vectors, we first specify a target number of representatives. We then partition the feature set into clusters by minimizing the within-cluster sum of squared Euclidean distances. This step yields a set of cluster assignments and associated centroids in feature space. Rather than using the centroids directly, we identify, for each cluster, the structure whose feature vector lies closest to the cluster centroid. This structure is the cluster medoid and is selected as the representative of that region of feature space. The final reduced set thus contains one medoid per cluster.

We also employ a density-based approach based on HDBSCAN. In the HDBSCAN framework, clusters are defined as regions of high local point density separated by areas of lower density.[20] Starting from a hierarchy of density-based clusterings constructed over a range of distance thresholds, HDBSCAN condenses this hierarchy and extracts a set of "stable" clusters that are robust to variations in the density parameter. Points that do not belong to any sufficiently stable region are labeled as noise. The method therefore simultaneously determines (i) the number of clusters, (ii) their membership, and (iii) which structures should be regarded as rare or outlying.

Within each HDBSCAN-identified cluster, we again select a single representative structure by choosing the configuration whose feature vector is closest to the cluster center in feature space. All points labeled as noise are retained individually, reflecting their potential importance as atypical or rare structural motifs. The resulting selection can be written as the union of all cluster representatives and all noise points. In contrast to K-means, the total number of selected structures is not fixed beforehand; instead, it emerges from the data through the interplay between the minimum cluster size, density thresholds, and the intrinsic structure of the feature distribution. This density-based scheme is particularly valuable when the structural landscape exhibits a mixture of common motifs and a small number of unique or weakly populated configurations. By automatically adjusting the number and shapes of clusters, HDBSCAN provides an adaptive, data-driven alternative to K-means that is less sensitive to manual tuning of the target set size.

### 2.4 High-throughput structure relaxation and stabilities evaluation

To further accelerate the exploration of the configurational space and introduce an explicit physical criterion into the selection pipeline, we employ pretrained uMLIPs for structure optimization. Such models, typically based on message-passing neural networks or related architectures, are trained on large and chemically diverse datasets of density-functional theory (DFT) energies and forces, and can therefore provide reliable predictions for a broad range of crystal structures at a fraction of the cost of direct DFT evaluations. In the present work, uMLIPs are used to perform geometry optimizations for large batches of candidate structures. The corresponding forces are obtained by uMLIPs and fed into a conventional local optimization scheme (e.g., quasi-Newton) to iteratively update the atomic positions and, when appropriate, relax the cell shape and volume. Each optimization is terminated when the maximal force component falls below a prescribed threshold or when a maximum number of steps is reached.

For each relaxed configuration, we extract approximate energetic and structural indicators. These quantities are subsequently used to define a physics-driven filtering step. Structures that fail to converge or lie far above the low-energy region, or show imaginary frequencies in phonon spectra are discarded. Conversely, low-energy and well-relaxed configurations are retained as promising candidates and passed on to more accurate DFT-level calculations. In this way, SRSS enable a multi-stage workflow that combines high-throughput structure generation, diversity-aware representative selection, and fast, physically informed relaxation, substantially reducing the number of structures that must be treated at the ab initio level while maintaining a broad and physically meaningful coverage of the structural landscape.

### 2.5 Computational Parameters

Mattersim-v1.0.0-1M was employed as the foundation model for structural relaxation, while Mattersim-v1.0.0-5M was utilized for phonon spectrum calculations in bulk materials.[21] For low-dimensional systems, the pre-trained Alex branch of the DPA-3 model was employed for structural optimization to ensure accurate treatment of surface and vacuum effects.[22, 27] Structural optimization, crystal structure processing, and symmetry determination were performed using the Atomic Simulation Environment (ASE),[28] Pymatgen,[29] and Spglib.[30] A symmetry tolerance of 0.02 Å was used for space group identification during the high-throughput generation and screening stages.

All first-principles calculations were performed within the framework of Density Functional Theory (DFT) using the Vienna Ab initio Simulation Package (VASP).[31-34] The Perdew-Burke-Ernzerhof (PBE) exchange-correlation functional within the Generalized Gradient Approximation (GGA) was adopted.[35] The plane-wave energy cutoff was set to 520 eV. Brillouin-zone integration was performed using Γ-centered k-point meshes with a spacing of $2\pi \times 0.03$ Å$^{-1}$.

## 3. Results & Discussions

For computational demonstration, we apply our SRSS to the following cases: 3D-SiC, 3D-BaPtAs, 2D-$NbSe_2$, and 1D-GaN. These systems were selected to validate the versatility and unbiased nature of the SRSS framework across diverse dimensionalities and chemical complexities. Specifically, SiC serves as a benchmark for polytypism in covalent bulk crystals, challenging the method to recover known stacking sequences while exploring unconventional symmetries beyond the common 3C and H phases.[36] The ternary compound BaPtAs represents a more complex multicomponent bulk system, testing the algorithm's ability to navigate a vast configurational space to identify dynamically competitive polymorphs alongside experimental ground states.[37] Moving to low dimensions, $NbSe_2$ is chosen to assess the discovery of stable 2D layered phases, notably aiming to uncover metastable structures with distinct electronic properties.[38-39] Finally, GaN provides a rigorous test for 1D nanotubular architectures, demonstrating the capability of SRSS to identify complex hollow structures purely from composition without relying on predefined rolling templates.[40] Collectively, this diverse set of benchmarks illustrates the robustness of SRSS in mapping the full landscape of crystal stability from simple binaries to complex low-dimensional nanostructures.

### 3.1 SiC semiconductors

Silicon carbide (SiC) is a classic example of polytypism in crystalline materials. Its most commonly studied phases include the cubic zincblende structure (3C) and a series of hexagonal (H) or rhombohedral (R) polytypes, such as 2H, 4H, and 15R.[36] These structures, characterized by similar tetrahedral bonding but different long-range stacking sequences, have dominated experimental and theoretical studies for decades. In contrast, SiC phases adopting distinct crystal symmetries beyond the conventional stacking family remain largely unexplored. Here, we leverage SRSS to explore the vast configurational space of SiC in an unbiased start to identify dynamically stable SiC polymorphs beyond the conventional set.

We systematically compared the performance of different descriptor-clustering strategies for the identification of phonon-stable SiC polymorphs from a generated dataset encompassing over 68,000 SiC structures across 229 (*P*1 was not considered) space groups. Two types of structural descriptors were employed: (i) low-dimensional, physically motivated geometric features (denoted as "simple"), and (ii) high-dimensional SOAP descriptors reduced via principal component analysis (PCA) to a compact subspace that retains 95% of the total variance. Each representation was paired with either *K*-means ($K = 30$ or 60) or HDBSCAN clustering.

Figure 1 illustrates the Uniform Manifold Approximation and Projection (UMAP) dimensionality reduction plots of structures selected from initial pool based on various combinations.[41] It is evident that simple descriptors achieve a relatively even coverage of the initial structural characteristics compared to the more complex SOAP descriptor. Moreover, as indicated in Figures 1a and 1d, alongside the statistical outcomes presented in Table 1, when employing the HDBSCAN method, the complexity of SOAP descriptor results in more challenging clustering of structures at a moderate set of minimum cluster size ( set as 3 for each space group). This leads to a significant increase in the number of structures retained, particularly those categorized as *rare-noisy*. Additionally, when applying the *K*-means method, structures selected under the simple descriptor exhibit a markedly more uniform coverage across the phase space relative to their SOAP counterparts, as demonstrated in Figures 1b, 1e, and 1c, 1f.

Figure 1g visualizes the evolutionary trajectories of space groups for the top 20 non-P1 symmetries using a Sankey diagram, based on the simple_km ($K$=60) subset. The flow reveals a distinct convergence trend: the most populated space groups after relaxation are *Fm*-3*m* (225), *Fd*-3*m* (227), and *Pm*-3*m* (221), all belonging to the cubic system. This indicates that high-symmetry cubic phases act as major attractors in the potential energy landscape of SiC. Tracing the flow backwards, the dominant *Fd*-3*m* phase primarily originates from precursors with cubic symmetry, such as *Fd*-3*m*, *Fd*-3, and $F4_132$, demonstrating a strong symmetry inheritance within the cubic family. Conversely, specific low-symmetry structures, such as those in space groups $P2_1$ (4) (Monoclinic) and $P3_2$ (145) (Trigonal), exhibit a *rigid* behavior where the space group remains invariant during relaxation. This heterogeneity in structural evolution, ranging from symmetry convergence to strict preservation validates the unbiased nature of SRSS. It confirms that our method can capture the diverse relaxation pathways of the system without forcing structures into a single prototype, balancing computational efficiency with configurational diversity.

To provide a straightforward assessment of the capability of different selection strategies in discovering polymorphs, we performed structural relaxation for all selected structures using uMLIP and extracted the 100 structures with the lowest energies. After removing similar structures, we computed their phonon spectra using the Mattersim, which was demonstrated to achieve a 95% accuracy in qualitatively predicting dynamical stability via phonon analysis. From this set, we quantified two key metrics: (i) the number of dynamically stable structures among the top 100 energetically ranked structures ($N_{ph_stable}$), and (ii) the number of distinct space groups represented within these dynamically stable structures ($N_{SG_ph_stable}$). The results are summarized in Table 1.

**Table 1**. The number of filtered crystals comparing different descriptor-clustering strategies. Nsel: Number of selected crystals from original structure pool. $N_{ph_stable}$: Number of dynamically stable structures among the top 100 energetically ranked structures after the removal of duplicates. $N_{SG_ph_stable}$: Number of different space groups in $N_{ph_stable}$.

| Groups | $N_{sel}$ | $N_{ph_stable}$ | $N_{SG_ph_stable}$ |
|---|---|---|---|
| simple_hd | 11786 (1137 + 10649) | 30 | 25 |
| SOAP_hd | 27118 (2091 + 25027) | 35 | 27 |
| simple_km ($K$ = 30) | 6846 | 33 | 29 |
| SOAP_km ($K$ = 30) | 6846 | 25 | 22 |
| simple_km ($K$ = 60) | 13686 | 33 | 30 |
| SOAP_km ($K$ = 60) | 13686 | 38 | 33 |

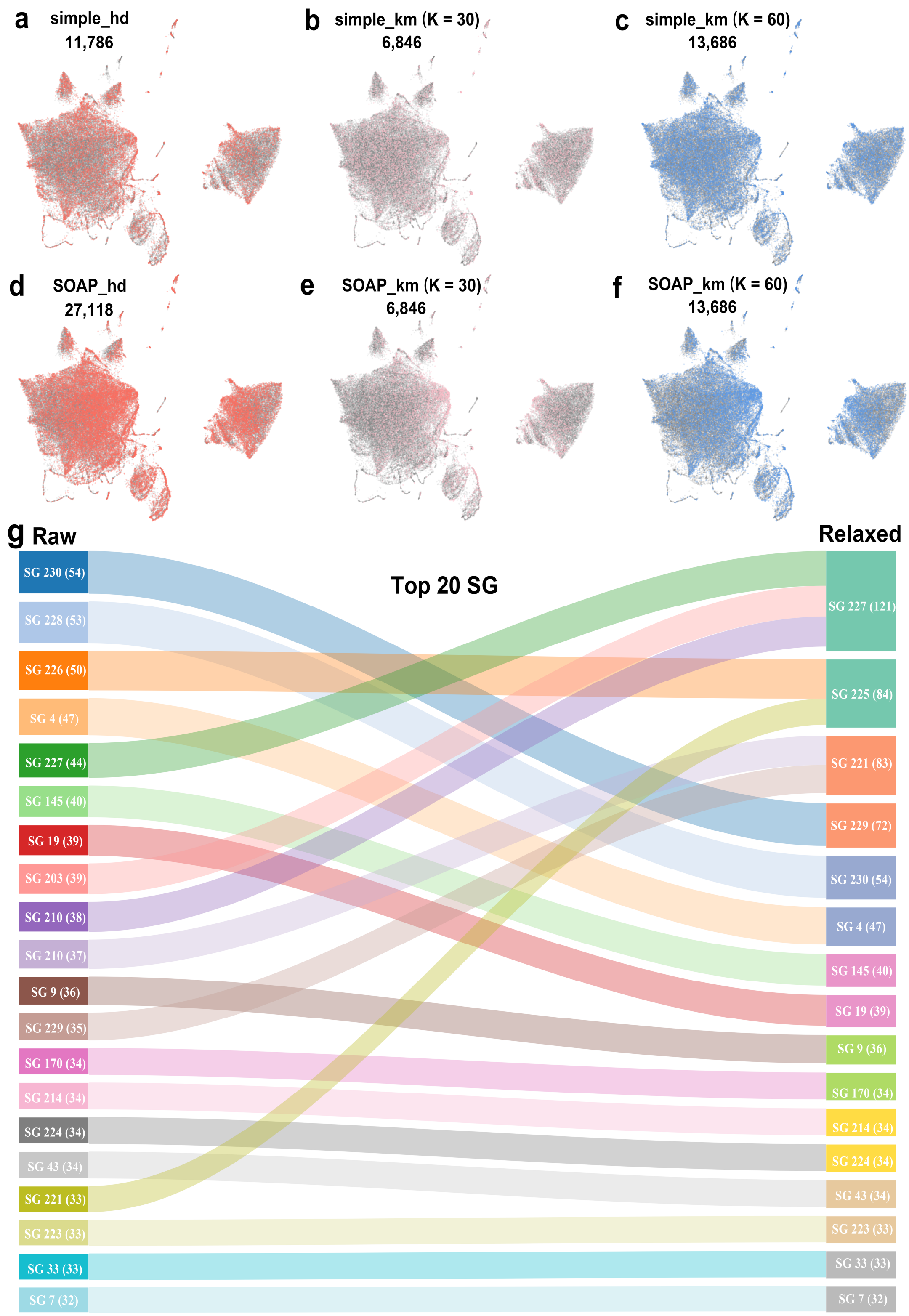

**Fig 1.** UMAP visualization of selected SiC crystals within the full set of raw structures, comparing different descriptor–clustering strategies. (a) simple geometric features with HDBSCAN, (b) simple geometric features with *K*-means ($K = 30$), (c) simple geometric features with *K*-means ($K = 60$); (d) SOAP descriptors with HDBSCAN; (e) SOAP descriptors with *K*-means ($K = 30$); and (f) SOAP descriptors with *K*-means ($K = 60$), (g) Sankey diagram illustrating the evolution of space group symmetries for the simple_km ($K$=60) subset before and after relaxation.

The results reveal that the performance gain of SOAP-based combinations is inconsistent compared to the simple descriptor. While two SOAP combinations show a marginal increase in the number of identified metastable structures (reaching a maximum of 38 with SOAP_km ($K$ = 30), the third combination shows no improvement. This indicates that the overall benefit is relatively limited. Notably, simple_km ($K$ = 30) identifies 33 metastable structures by screening only 6,846 raw structures, achieving the highest screening efficiency among all tested combinations. Furthermore, these 33 structures span 29 distinct space groups, corresponding to a diversity ratio of 88 %, which is on par with the best SOAP-based result ( SOAP_km ($K$ = 60): 33 out of 38, or 87%).These observations again indicate that the use of more complex descriptors such as SOAP provides only limited gains in this screening context. In contrast, the combination of simple, interpretable descriptors with $K$-means clustering offers a highly efficient and robust pathway for initial polymorph discovery, achieving near-optimal performance with significantly reduced computational overhead. Consequently, for large-scale exploratory studies where rapid and reliable identification of stable candidates is prioritized, simple descriptors coupled with $K$-means are sufficient and practically advantageous in SRSS workflow.

Initially, for each combination investigated, both the *H* phase (*P*63*mc*) and the 3C phase (*F*-43*m*) with various stackings matching known experimental results were identified. It is noteworthy that the *R* phase structure was not discovered within the screening outcomes under the current limitation of atomic numbers. Given that the unit cells of the *R* phase structures tend to contain a larger number of atoms, exploring these may require a more specifically designed initial structure generation for only larger unit cells, highlighting the complexity of the SiC system.

Subsequently, the structures identified under different combinations were further examined using DFT calculations. An investigation into previously reported structures revealed that four of our predicted Si-C polymorphs align with prior computational predictions, having space group numbers *Pbca*, *Pbcn*, *Pm*, and *Pnnm*, respectively. Additionally, several unreported structures were predicted, including *Aem*2, $P4_32_12$, *I*4*cm*, etc. Three representative findings are displayed in Figure 2: the $P4_2/mnm$ phase presenting a Euclidean tiling pattern from a top view, the *Pm*-3*n* phase featuring truncated octahedral cages of SiC, and the complex *Ccc*2 phase with a 40-atom unit cell. As a validation step, we performed phonon dispersion calculations at the PBE level of DFT for these structures, also confirming their dynamical stability. The band calculations indicate these new phases are also semiconducting, as shown in **Fig S1**.

The SiC screening results here not only reveal the vast polymorphic landscape of the Si-C system but also demonstrate that the final candidate structures are not confined to the commonly observed 3C or *H* phases. This underscores the unbiased nature of the SRSS (Structure Representation and Selection Strategy) methodology, which enables exploration beyond conventional structural motifs.

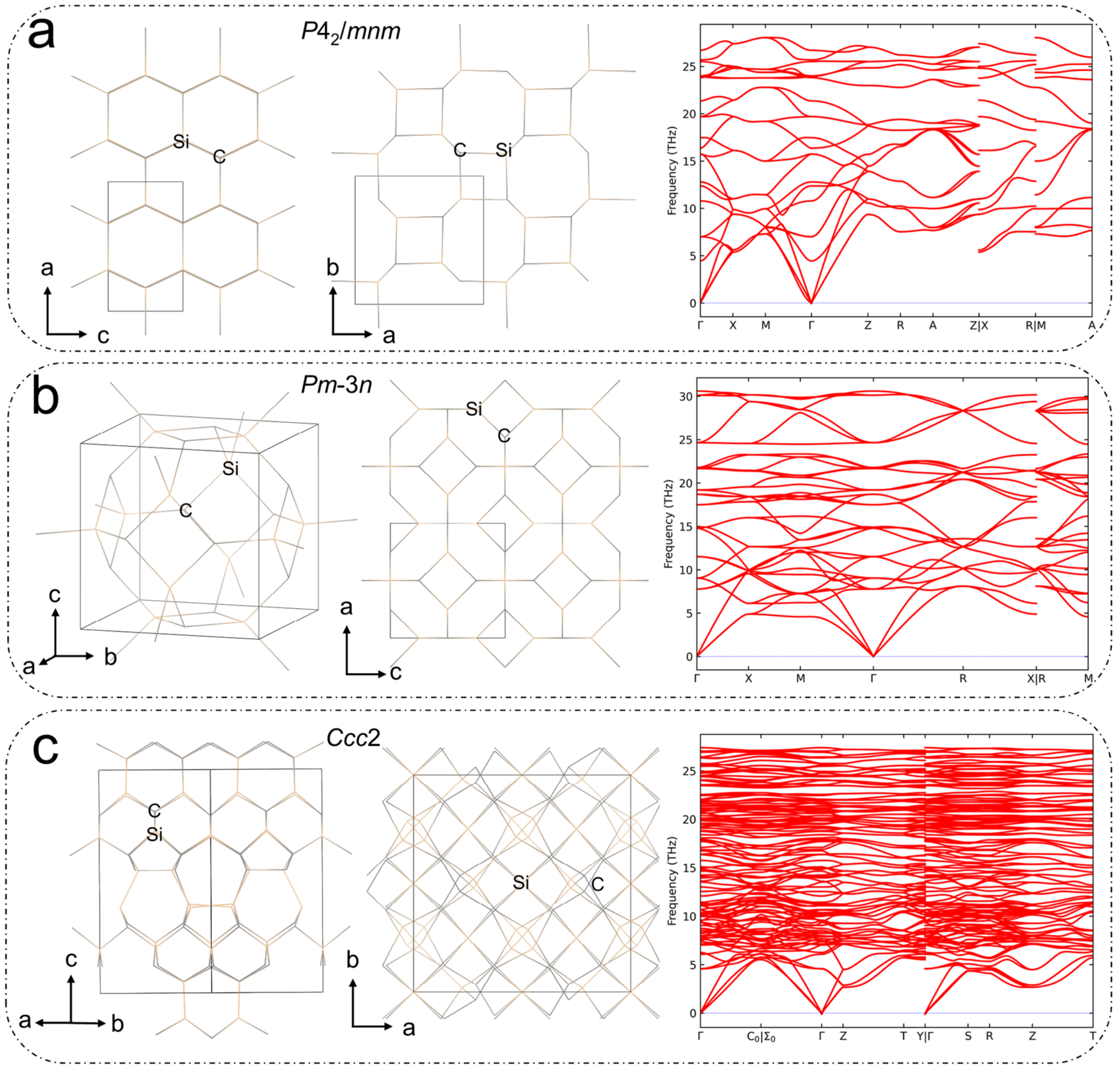

**Fig 2.** Schematic diagram of crystal structure and phonon spectrum calculated by DFT-PBE for a) $P4_2/mnm$, b) *Pm*-3*n* and c) *Ccc*2 SiC polymorph phases.

### 3.2 Ternary Compound BaPtAs

Subsequently, we further investigated the screening capability of the SRSS method for more complex bulk systems. In this context, an equiatomic ternary compound, BaPtAs, which has been experimentally synthesized into three distinct crystal structures, was chosen as the subject. Figure 3a illustrates the selection scenario for this system. Initially, a total of 64,200 structures were randomly generated, following which the *K*-means method (retaining 60 structures per space group) was employed to sieve out 12,840 structures. Post the initial filtering, all selected structures underwent structural optimization, from which a few structures that exhibited phonon stability and lowest energy at the uMLIP precision level were retained. Furthering the analysis, DFT accuracy energy calculations were performed on these optimized structures, and based on their $E_{hull}$ values calculated at DFT precision, only those with an $E_{hull}$ value less than 0.05 eV/atom were kept. The configurations of these structures are presented in Figure 3b, while their $E_{hull}$ calculation outcomes are depicted in Figure 3c.

Kudo et al. experimentally reported three crystal structures of BaPtAs,[37] namely the YPtAs-type, SrPtSb-type, and LaIrSi-type phases, corresponding to space groups $P6_3/mmc$, $P$-6$m$2, and $P2_13$, respectively, as shown in Figure 3b. Among them, the $P2_13$ structure exhibits an $E_{hull}$ value close to zero, while the $P$-6$m$2 and $P6_3/mmc$ phases have $E_{hull}$ values of approximately 0.02 eV/atom. As shown in Figure 3c, the fourth lowest $E_{hull}$ structure belongs to the $I4_1md$ space group, a phase that was previously analyzed and discussed in our earlier work and confirmed to possess a phonon spectrum free of imaginary frequencies. Beyond these experimentally known polymorphs and the theoretically explored $I4_1md$ phase, the SRSS approach identified two additional low-energy candidates with $E_{hull}$ < 0.05 eV/atom, belonging to the *Pbca* and $P2_1/c$ space groups, respectively. Although we have already calculated the phonon spectra using uMLIP, we performed further DFT-based phonon calculations to rigorously verify their dynamical stability. The results presented in Figures 3d and 3e confirm the absence of imaginary modes, indicating that both structures are dynamically stable and thus viable polymorphs within the BaPtAs system. Collectively, the screening results for the BaPtAs system demonstrate that the SRSS method balances simplicity and performance in identifying potentially stable polymorphs in complex multicomponent systems.

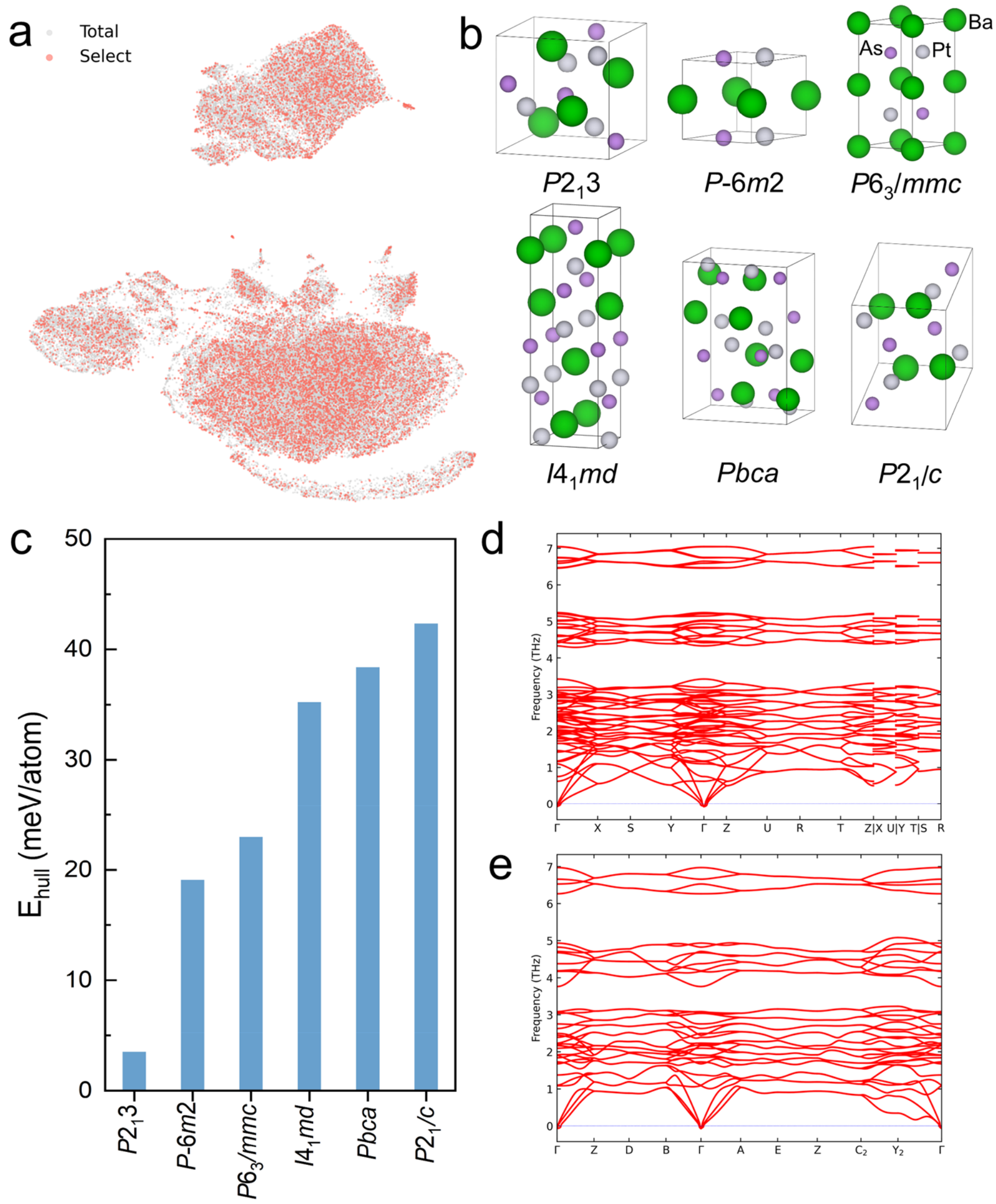

**Fig 3.** a) UMAP visualization of selected BaPtAs crystals within the full set of raw structures. b) Schematic representations of the crystal structures of identified BaPtAs polymorphs. c) $E_{hull}$ results of BaPtAs polymorphs computed at the DFT level. d,e) DFT-calculated phonon dispersion spectra for d) *Pbca* and e) $P2_1/c$ polymorphs, respectively.

### 3.3 Low dimensional compounds

To assess the applicability of the SRSS method to low-dimensional crystals, we performed a structure prediction study on the well-known 2D-$NbSe_2$. Following high-throughput structure generation and feature extraction, we screened a total of 7,900 candidate structures. Taking into account the suitability of dimensionality-reduced models, structural relaxation was carried out using the Alex branch of the DPA-3 model.[22] After excluding one-dimensional atomic-chain configurations, in addition to the well-established 1*H* and 1*T* phases, we identified a previously unreported orthorhombic polymorph, which was designated as the 1*O*-$NbSe_2$. As illustrated in Figure 4a, this 2D 1*O*-$NbSe_2$ structure is composed of periodically tiled Nb-Se eight-membered rings and distorted Nb-Se quadrilaterals.

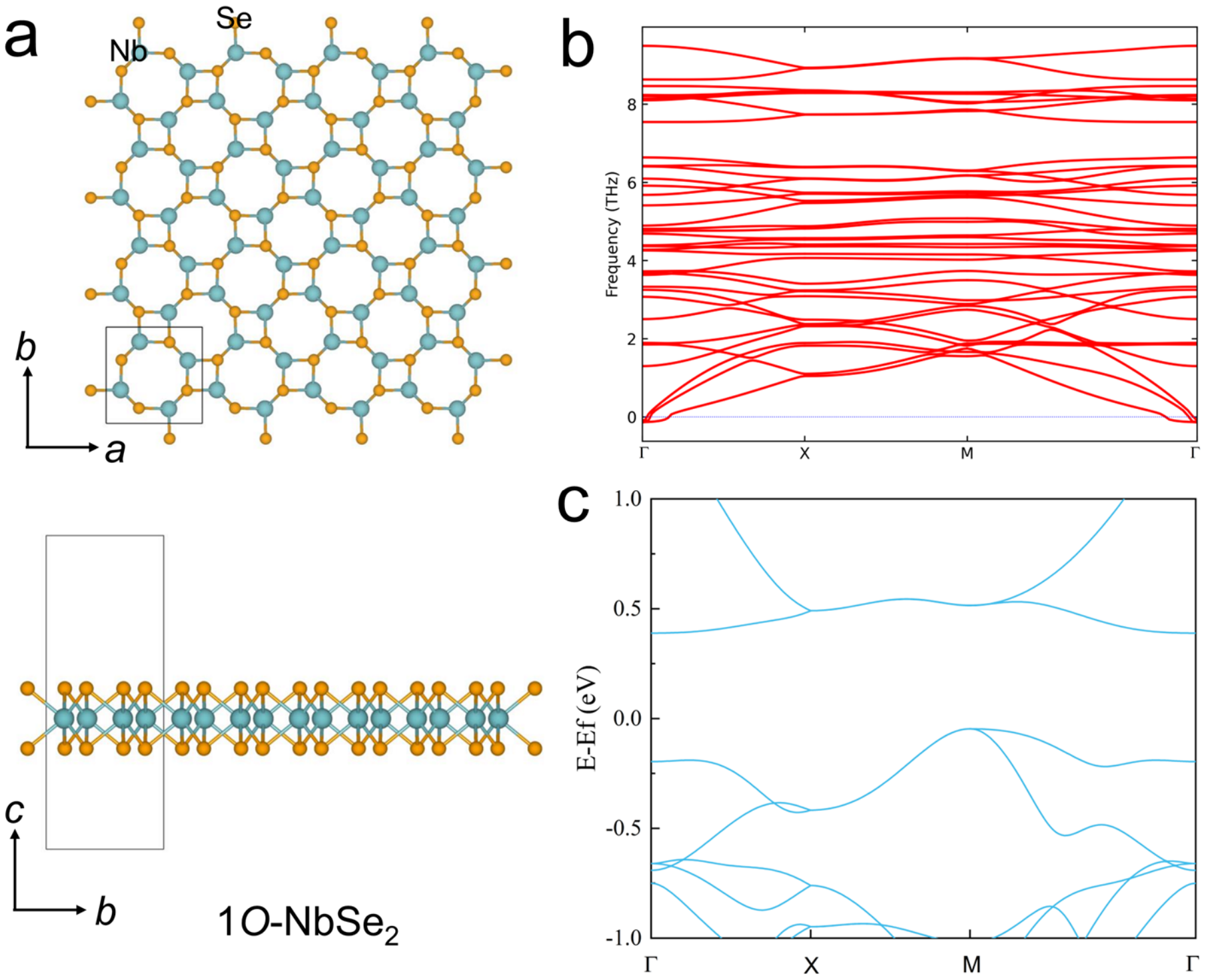

**Fig 4.** a) Schematic representations of the crystal structure of predicted 1*O*-$NbSe_2$. b) DFT-calculated phonon dispersion and c) energy band.

DFT calculations reveal that its total energy lies 0.18 eV/atom above that of the most stable 2D 1*H*-$NbSe_2$ polymorph. Nevertheless, the phonon dispersion spectrum computed at the DFT level (Figure 4b) confirms the dynamical stability of the 1*O* phase, as it exhibits no imaginary frequencies across the Brillouin zone. Furthermore, the electronic band structure of 1*O*-$NbSe_2$, shown in Figure 4c, reveals a non-negligible band gap, an intriguing contrast to the metallic character of both the 1*H* and 1*T* phases. To our knowledge, no semiconducting polymorph of 2D $NbSe_2$ has been reported to date. The discovery of this semiconductor 1*O* phase underscores the capability of the SRSS methodology to explore and identify stable configurations even in low-dimensional systems.

To further evaluate the applicability of the SRSS method to 1D nanotubular systems, we assessed its capability in screening 1D GaN system. Notably, we excluded 1D atomic chains from consideration, as they typically occupy a much simpler region of the configurational space and are more readily sampled by random structure generators. In contrast, hollow nanotubes, requiring precise curvature, closure, and bond saturation represent a structurally complex motif that is significantly underrepresented in unbiased random searches, thereby posing a greater challenge for discovery algorithms Specifically, we selected the binary compound GaN as a representative system and initiated our search from the complete set of rod groups to explore the ability of SRSS to identify viable GaN nanotube configurations without relying on preconceived structural templates. As shown in Figure 5, our approach successfully uncovered three distinct nanotube structures: (3,3)-armchair GaN, (4,4)-armchair GaN, and (6,0)-zigzag GaN, following standard nanotube nomenclature. Electronic structure calculations reveal that both the (3,3)-armchair and (6,0)-zigzag nanotubes exhibit direct band gaps, whereas the (4,4)-armchair variant displays an indirect band gap, highlighting the rich electronic diversity accessible within this class of 1D materials.

Critically, unlike conventional nanotube construction strategies that begin with rolled-up 2D sheets or other heuristic templates,[42] our protocol required only elemental composition and stoichiometry as input. In other words, these nanotube structures were identified from a vast pool of completely unconstrained candidate configurations, without any prior structural assumptions. Although the known configurational space of GaN nanotubes is extensive and our current search did not exhaustively sample all possible chiralities or diameters,[40] the successful identification of three structurally and electronically distinct nanotubes, spanning both armchair and zigzag families demonstrates the robustness and transferability of the SRSS framework to complex low-dimensional systems, including challenging 1D nanotubular architectures.

The polymorph screening results across the aforementioned systems, spanning bulk crystals, 2D layered materials, and 1D nanotubular architectures, collectively demonstrate that SRSS is a robust and versatile framework for unbiased discovery of thermodynamically competitive and dynamically stable phases. Nevertheless, it is important to acknowledge that the reliability and structural diversity of the candidates identified by SRSS are inherently contingent upon the accuracy of the uMLIP employed during the relaxation stage. While the current uMLIP enables high-throughput screening of tens of thousands of configurations at minimal computational cost, its training dataset is predominantly biased toward known stable phases. Consequently, it may exhibit reduced fidelity for highly distorted geometries, metastable motifs far from equilibrium, or systems governed by strong electronic correlations. The ongoing development of more accurate and generalizable MLIPs, such as those based on equivariant graph neural networks or augmented by active learning strategies, is expected to substantially enhance SRSS's capability to capture high-energy metastable states and even transient intermediates, all while preserving its high-throughput nature.

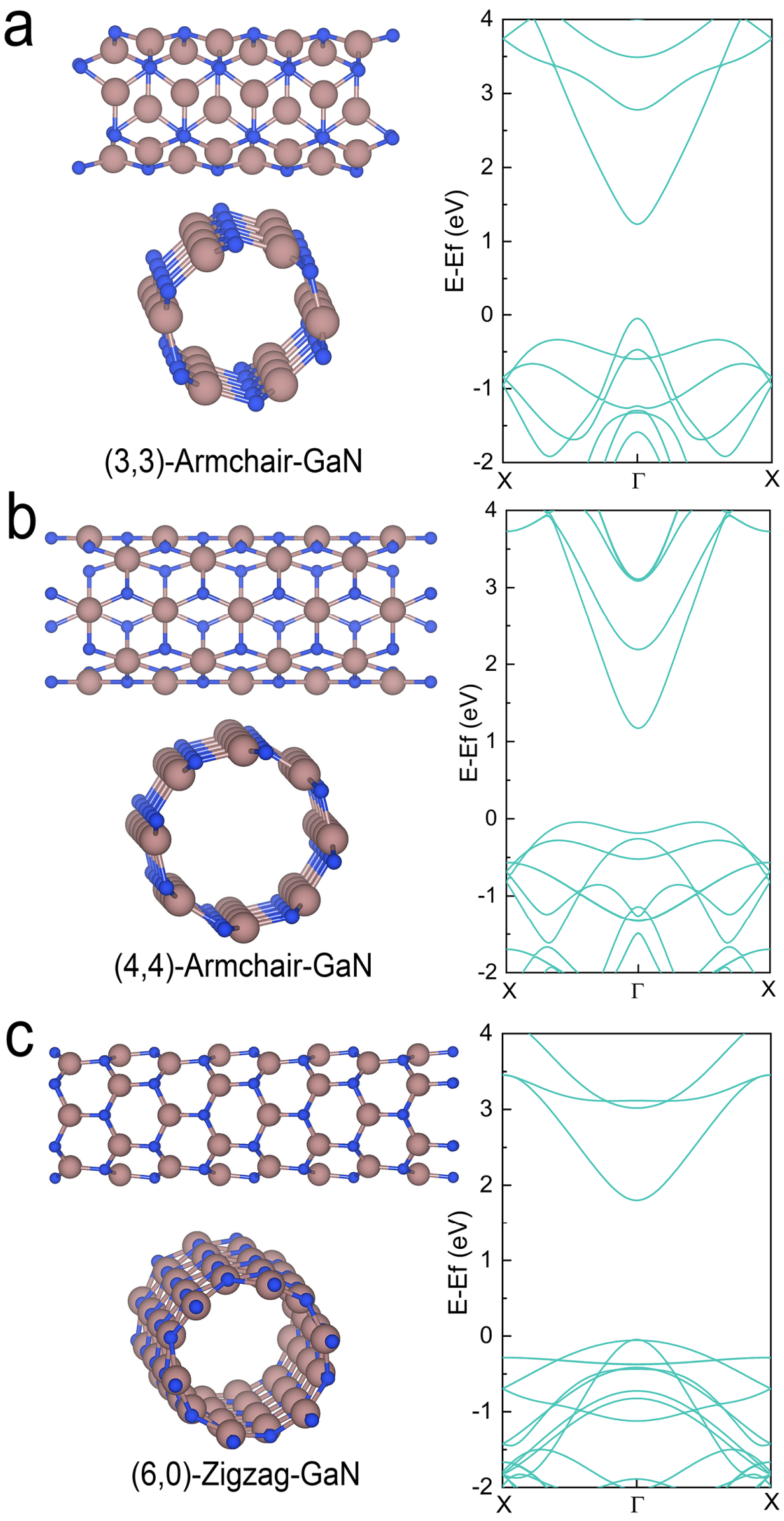

**Fig 5.** Schematic representations of the crystal structure and energy band of GaN nanotubes

predicted in this work: a) (3,3)-Armchair GaN, b) (4,4)-Armchair GaN and (6,0)-Zigzag GaN.

Notably, SRSS achieves high throughput using only standard CPU resources, with no need for specialized hardware such as GPUs, making it accessible even on entry-level workstations or laptops. For instance, in the BaPtAs case study, the generation of 64,200 symmetry-constrained candidate structures required only ~20 minutes on a single 64-core CPU node, while the subsequent diversity-based selection took less than 3 minutes. The relaxation of the selected 12,840 structures using uMLIP was completed within 75 minutes, thanks to efficient parallelization across space groups. In principle, the entire workflow can be executed on a standard laptop with a quad-core processor, making SRSS particularly suitable for exploratory studies in resource-limited settings. Given that the relaxation step dominates the computational cost and that this cost scales directly with the complexity of the underlying uMLIP, the continued advancement of model compression, distillation, and fine-tuning techniques will further reduce the barrier to exploring increasingly complex chemical spaces.[43-44]

# Conclusion

In summary, this work presents SRSS as a powerful paradigm for the unbiased discovery of crystal polymorphs across 3D bulk, 2D layered, and 1D nanotubular architectures. By decoupling structure generation from energetic priors and leveraging symmetry groups to ensure comprehensive coverage, SRSS overcomes the inherent sampling biases of conventional methods, successfully identifying both thermodynamically competitive ground states and kinetically accessible metastable phases that exhibit unique electronic and topological properties. The application of SRSS to 3D-SiC, 3D-BaPtAs, 2D-$NbSe_2$, and 1D-GaN not only validates its efficacy in recovering known experimental structures but, more significantly, expands the known phase space of these materials with novel candidates such as the semiconducting 1O-$NbSe_2$ and diverse GaN nanotubes. Furthermore, the integration of uMLIPs for rapid relaxation enables this exhaustive search to be performed with remarkable computational efficiency, requiring only modest CPU resources and eliminating the need for specialized hardware. While the current fidelity of discovered structures depends on the generalization capability of existing uMLIPs, the framework is inherently future-proof: as next-generation potentials trained on broader datasets emerge, SRSS will seamlessly integrate these advances to probe even more complex, strongly correlated, or highly distorted regimes. Ultimately, SRSS democratizes high-throughput crystal exploration, offering a practical, physics-driven tool for accelerating materials discovery in both academic and resource-constrained environments.

## Author's contributions

**Jiexi Song**: Conceptualization, calculation, data analysis, preparation of paper. **Diwei Shi**: discussion, paper editing and review. **Aixian She**: discussion. **Chongde Cao**: paper editing and review, funding acquisition. **Fengyuan Xuan**: discussion, paper editing and review, funding acquisition.

## Data availability

The crystal structure files supporting the findings of this study are available in the Zenodo repository with the DOI: 10.5281/zenodo.19446966.

## Conflicts of interest

There are no conflicts of interest to declare.

## Acknowledgements

This work was supported in part by the Basic Research Program of Jiangsu (Grant No. BK20240395), the National Natural Science Foundation of China (52271037), the Shaanxi Provincial Natural Science Fundamental Research Program, China (2025SYS-SYSZD-098), the Jiangsu Funding Program for Excellent Postdoctoral Talent (Grant No. 2025ZB701) and Opening Grant of Zhejiang Key Laboratory of Data-Driven High-Safety Energy Materials and Applications (OG2024008). Calculations were performed on Sugon HPC clusters equipped with HYGON X86 32-core processors (2.5 GHz) and at the Beijing Super Cloud Computing Center.